\documentclass[leqno,openbib]{article}
\usepackage{float}
\usepackage{lineno}
\usepackage{url}
\usepackage[utf8]{inputenc}
\usepackage{booktabs}
\usepackage{indentfirst}
\usepackage{amsfonts}
\usepackage{ntheorem}
\usepackage{amsmath}
\usepackage[singlespacing]{setspace}
\usepackage{sectsty}
\usepackage[norule,bottom]{footmisc}
\usepackage[justification=centering,textfont={sc},labelfont={rm}]{caption}
\usepackage{varioref}
\usepackage{natbib}
\setcitestyle{authoryear,open={(},close={)}}
\usepackage[labelfont=bf]{caption}
\usepackage [autostyle, english = american]{csquotes}
\MakeOuterQuote{"}
\usepackage{graphicx}

\theoremindent\parindent
\makeatletter    
\renewtheoremstyle{plain}{\item[\hskip\labelsep\hskip-\parindent \theorem@headerfont ##1\ ##2\theorem@separator]}{\item[\hskip\labelsep\hskip-\parindent \theorem@headerfont ##1\ ##2\ (##3)\theorem@separator]}
\makeatother
\theoremheaderfont{\scshape}
\theorembodyfont{\rmfamily}
\theoremseparator{.}

\sectionfont{\normalfont\scshape\centering}

\subsectionfont{\itshape}
\makeatletter
\def\@biblabel#1{\hspace*{-\labelsep}}

\makeatother
\makeatletter
\renewcommand\@makefnmark{\mbox{\textsuperscript{\normalfont\@thefnmark}}}
\renewcommand\@makefntext[1]{\indent\makebox[2.5em][r]{\@thefnmark.\,}#1}
\if@titlepage
  \renewenvironment{abstract}{      \titlepage
      \null\vfil
      \@beginparpenalty\@lowpenalty
      \begin{center}        \@endparpenalty\@M
      \end{center}}     {\par\vfil\null\endtitlepage}
\else
  \renewenvironment{abstract}{      \if@twocolumn
      \else
        \small
        \begin{center}        \end{center}      \fi}
      {\if@twocolumn\else\endquotation\fi}
\fi
\renewcommand\thetable{\@Roman\c@table}

\renewcommand\thefigure{\@Roman\c@figure}

\makeatother
\labelformat{equation}{(#1)}

  \newenvironment{hypothesis}{
  	\itshape
  	\leftskip=\parindent \rightskip=\parindent
  	\noindent\ignorespaces}

\begin{document}

\title{The Wisdom of a Kalman Crowd}
\author{
\begin{tabular}{c}
\textsc{Ulrik W. Nash} \\ 
\textsc{Syddansk Universitet} \\ 
\textsc{5230 Odense M, Denmark.} \\
\textsc{Email: uwn@sam.sdu.dk} \\
\end{tabular}
\textsc{}}
\maketitle

\begin{abstract}
The Kalman Filter has been called one of the greatest inventions in statistics during the 20th century. Its purpose is to measure the state of a system by processing the noisy data received from different electronic sensors. In comparison, a useful resource for managers in their effort to make the right decisions is the wisdom of crowds. This phenomenon allows managers to combine judgments by different employees to get estimates that are often more accurate and reliable than estimates, which managers produce alone. Since harnessing the collective intelligence of employees, and filtering signals from multiple noisy sensors appear related, we looked at the possibility of using the Kalman Filter on estimates by people. Our predictions suggest, and our findings based on the Survey of Professional Forecasters reveal, that the Kalman Filter can help managers solve their decision-making problems by giving them stronger signals before they choose. Indeed, when used on a subset of forecasters identified by the Contribution Weighted Model, the Kalman Filter beat that rule clearly, across all the forecasting horizons in the survey.\\
~\\
\textbf{JEL-codes:}  C53, D80, D82, D84, D87.
\end{abstract}

\section{Introduction}
\noindent When managers decide what to do, they often base their choices on forecasts of future magnitudes. Examples are easy to find. Before deciding whom to hire, the manager estimates future productivity, before setting the path of R\&D, the manager estimates future returns, before diversifying, the manager estimates future demand, and so the list goes on. In each case, estimates with lower errors are better. Unfortunately, to reduce error is not trivial, and managers often make the wrong choice because of their imprecision.

When it comes to forecasting errors, imperfections of human cognition play an essential role. Although our cognitive system is an outstanding device for measuring the present, it has more difficulty estimating things yet to occur. Estimating the weight of milk, the distance to the carton, and the amount of effort needed to serve a glass is unproblematic for most people, but when asked to estimate the amount of milk they will spill during the next year despite their sensory and motor skill, most people feel uncertain and make sizable mistakes. As the forecast horizon reaches further into the future, the faults of human cognition create an uncertain relation between what we think will happen, and what does.

Scholars have long known \citep{Galton1907} that while the estimates we make alone are often subject to glaring errors, we can combine them with estimates made by others to form an aggregate that improves both accuracy and precision. Scholars have called this phenomenon many things, from Vox Populi \citep{Galton1907} to Rational Expectations \citep{Muth1961}, to the Many Wrongs Principle \citep{Simons2004}, but most recently \cite{Surowiecki2004} called it the Wisdom of the Crowds, and that name has stuck.

There are many ways to fuse estimates from different sources and researchers have produced many rules for doing that. These include both formal rules where people never interact and social rules where they do. The most famous rule is the simple average or the Equal-Weight Model (EWM). However, other common rules include designs for social interaction like the Delphi method by the RAND Corporation \citep{Dalkey1963}, Social Judgment Analysis \citep{Hammond1965,Dhami2008} based on cue learning psychology \citep{Tolman1935}, and anonymous trading in prediction markets \citep{Wolfers2004} used first by the University of Iowa. Other rules grant access to information about prior performances \citep{King2012}, while some combine estimates by subsets of people and ignore those made by others \citep{Budescu2015, Mannes2014}. The latest rules in that last category have done particularly well \citep{Budescu2015, Chen2016}. More specifically, the so-called Contribution Weighted Model (CWM) has beaten many other rules not only because it has an algorithm for finding subsets of strong forecasters, but also because it puts different weights on their estimates using relative measures \citep{Budescu2015, Chen2016} of how they performed in the past.

It might surprise that when estimates are independent and unbiased, and the aim of combining estimates is to minimize the uncertainty and bias about the truth, and when we know what uncertainty to expect from each source of estimates, then none of the above rules are best. That is a matter of fact because the best rule, under those conditions, appeared sixty years ago \citep{Kalman1960} and turned out to be one of the greatest achievements in statistics during the 20th century \citep{Grewal2008}. This Bayesian rule not only guided Armstrong to the moon and back \citep{McGee1985} but now supports almost any technology that needs to estimate the state of the world from noisy data, including GPS and inertial guidance systems in planes. The rule in question is the Kalman Filter (KF).

Why researchers have developed so many rules to harness the wisdom of crowds without giving much attention to the KF, is a good question. One reason might be that engineers use the KF to combine estimates by electronic sensors, and electronic sensors seem very different from human perception. As such, the possibility of using the KF on human magnitude estimates has perhaps just been unclear. On the other hand, psychophysicists \citep{Luce1972} have long argued that humans are, among other things, measuring devices, which suggests the issue goes beyond perceived relevance. Indeed, the other reason may be actual relevance, given what we can know about the cognitive system. After all, Electronic sensors are specialized to domains, and we can discover their uncertainty through intensive tests and calibrations. In contrast, our knowledge about the source of human judgment (i.e., the brain) is vaguer, and we cannot run the same tests on people. Therefore, since the KF assumes good knowledge about the uncertainty produced by the source of measurements, researchers may have developed the many alternative rules because they suppose the KF   harnesses the wisdom of crowds poorly in practice. Nevertheless, we do not know with any certainty because few studies have tried to find out.

In this paper, we clarify things by moving the KF from engineering to management so that we can match it against the EWM and the CWR. We do this using a model of judgment, which appeared recently in mathematical psychology \citep{Nash2017}. The model lets us implement the KF on estimates by humans while offering a theory for the source of uncertainty in such estimates. The model also lets us ask at what point our knowledge of the uncertainty in estimates is so vague that the EWM and the CWM will beat the KF, and we derive precise answers to this question. We then test our prediction using sixty surveys of professional forecasters about economic variables over five forecasting horizons. 

Based on our findings we conclude, as Budescu and Chen recently did \citep{Budescu2015}, that the main attraction of the CWM is not in assigning optimal weight to estimates, but in identifying individuals with superior judgment ability. To that statement, we add that when the CWM has identified these judges, one ought to hand the job of assigning weights to the KF.

\section{The Kalman Filter}
\noindent A butcher and a farmer meet at a country market. The farmer has an ox that he wants to sell, and the butcher is interested in buying. Now, they must agree on a price.

There is a straightforward link between what oxen weigh, and what oxen cost in the market, so the butcher and the farmer must weigh the ox to reach an agreement. Regrettably, someone has broken the market's official scale. Both the farmer and the butcher, however, have scales of their own, but these are imperfect. Consequently, the relationship between their output and what the ox weighs is uncertain.

How should the butcher and the farmer use their scales to minimize the mean squared error (MSE) to the correct price? Assuming the measurements produced by the scales are probabilistic and independent, and assuming the degrees of uncertainty in the scales are known, then the KF offers the answer.

To see this, let us first consider the definition of MSE by using the butcher's scale as an example. Denoting the butcher's scale by subscript 1, the definition of MSE is

\begin{equation}
\label{eq:1}
MSE_1 = \sigma_1^2+(X-\mu_1)^2
\end{equation}

\noindent where $\sigma_1^2$ is the uncertainty of the butcher's scale, captured by the variance of the estimates it produces, $\mu_1$ is the mean estimate produced by the scale, and $X$ is the actual weight of the ox. 

From equation \ref{eq:1} we notice two things. First, $X - \mu_1$ captures how biased the butcher's scale is, and second, when $X = \mu_1$ the butcher's scale is unbiased, such that $MSE_1 = \sigma_1^2$.

Now let us introduce the farmer's scale, and denote it by subscript 2. The problem of how to use the two scales to minimize MSE amounts to working out how much weight to place on the estimates produced by each scale to obtain an aggregate estimate that minimizes MSE. We denote the weight placed on estimates produced by the butcher's scale as $w_1$ and denote the weight placed on estimates produced by the farmer's scale as $w_2$. Then we assume weights are best placed linearly, such that MSE of the aggregate estimate is

\begin{equation}
\label{eq:2}
MSE_{1,2} = (\sigma_1^2 w_1^2 + \sigma_2^2 w_2^2) + ((w_1 \mu_1 + w_2 \mu_2) - X)^2
\end{equation}

Next, we will assume that any weight not placed on estimates produced by the butcher's scale will instead be placed on the farmer's scales, thus establishing a weighted average. Accordingly, we can substitute $w_2$ with $1 - w_1$ and re-state equation \ref{eq:2} as

\begin{equation}
\label{eq:3}
MSE_{1,2} = (\sigma_1^2 w_1^2 + \sigma_2^2 (1 - w_1)^2) + ((w_1 \mu_1 + (1 - w_1) \mu_2) - X)^2.
\end{equation}

From here, we simply differentiate equation \ref{eq:3} with respect to $w_1$, set the resulting equation to zero, and solve for $w_1$. Performing this procedure (and verifying the second derivative is positive) we obtain the optimal weights to place on estimates by each scale

\begin{equation}
\label{eq:4}
w_1^* = \frac{\sigma _2^2+\left(\mu _1-\mu _2\right) \left(X-\mu _2\right)}{\left(\mu _1-\mu _2\right){}^2+\sigma _1^2+\sigma _2^2}.
\end{equation}

and

\begin{equation}
\label{eq:5}
w_2^* = 1 - \frac{\sigma _2^2+\left(\mu _1-\mu _2\right) \left(X-\mu _2\right)}{\left(\mu _1-\mu _2\right){}^2+\sigma _1^2+\sigma _2^2}.
\end{equation}

From equations \ref{eq:4} and \ref{eq:5} we notice the intuitive result that more weight should be placed on the estimate produced by the butcher's scale when the uncertainty of that scale is smaller compared to the uncertainty of the farmer's scale, holding the levels of bias constant. Accordingly, the weight associated with a particular scale is not in general fixed, but depends on its uncertainty relative the uncertainty of the second scale, and vice versa. That becomes clearer as we now introduce the assumption by \cite{Kalman1960} that sensors provide unbiased estimates. In this case, equations \ref{eq:4} and \ref{eq:5} are reduced to

\begin{equation}
\label{eq:6}
w_1^* = \frac{\sigma _2^2}{\sigma _1^2+\sigma _2^2}.
\end{equation}

\noindent and

\begin{equation}
\label{eq:7}
w_2^* = 1 - \frac{\sigma _2^2}{\sigma _1^2+\sigma _2^2}.
\end{equation}

\noindent which are commonly referred to as the Kalman Gains.

\subsection{Verification by the a third party:} 
\noindent The butcher and the farmer now have what they need to reach an agreement on price. What they must do is take their estimates, $x_1$ and $x_2$, weigh them by $w_1$ and $w_2$ respectively, sum what they obtain, and convert this sum to a price according to the market's rule.

However, suppose a market official wants to verify that everything is in good order. To do that, the official will use a third, unofficial scale to update the butcher's and the farmer's combined estimate. The aim is not only to reduce MSE further using new evidence but also to keep MSE minimized.

While the official could combine the estimates of two or more sensors at once and obtain an optimal result, part of the beauty of the KF is that it offers the possibility to combine estimates one by one, while remaining optimal at every step.

To see that, we insert the Kalman Gains into equation \ref{eq:2} and note the right side of the equation falls away. Equation \ref{eq:2} hence simplifies to

\begin{equation}
\label{eq:8}
MSE_{KF} = \frac{\sigma _1^2 \sigma _2^2}{\sigma _1^2+\sigma _2^2}.
\end{equation}

Because the butcher's scale and the farmer's scale are now unbiased, equation \ref{eq:8} equals the uncertainty of their combined estimate. At this point it is common in practice to assume that uncertainty is described not only by white noise, but Gaussian white noise. The reasons for that assumption, besides often being realistic for electronic sensors, is that engineers can thereby use the Gaussian distribution's rare property of closure under multiplication. Specifically, due to this property, we can express equation \ref{eq:8} simply as $\sigma _{KF}^2$, where this constant is the variance of another Gaussian distribution. As such, the combined estimate becomes no different, mathematically, from the butcher's estimate that started the process. Accordingly, we have reset the process computationally, while keeping it optimal. We can easily add more estimates using equations \ref{eq:6} and \ref{eq:7} without inflating the computational demand on the system that implements the filter. Moreover, as we do that, our approach to filtering the signal from noise remains the best of any conceivable form \citep{Maybeck1979}. Indeed, even if the Gaussian assumption is removed, the KF is the best (minimum error variance) filter out of the class of linear unbiased filters \citep{Maybeck1979}.   

Given the constraints of computers at the time, the reader can now appreciate why the KF became so crucial for Project Apollo. However, the reader must also realize the above exposition only revealed the essential part of the KF. When the KF directs navigation systems, for example, it must account for sensor readings received at different points in time. Consequently, since significant changes occur between readings, the dynamic KF accounts for those changes by modeling them. For example, navigation systems incorporate Newton's equations of motion. The static KF suffices, however, when we can reasonably assume that estimates are made simultaneously.

\section{The Augmented Quincunx}
\noindent Suppose the official scale was the only proper scale in the market. If the butcher and farmer both insist on trading, they would need to measure the weight of the ox using just their cognitive system. Furthermore, if the official wanted to verify that everything was in good order, he would need to do the same. Nevertheless, the KF is relevant in this setting too, and the Augmented Quincunx (AQ) model of judgment can be used to appreciate why that is.

The AQ recently \citep{Nash2017} joined the family of sequential sampling models. These models do an excellent job predicting observed patterns relating to choice \citep{Forstmann2016}. Moreover, they work at the algorithmic level \citep{Marr1982}, which means they not only predict the outcomes of thinking but also propose reasons for those outcomes. Furthermore, they have solid neurophysiological foundations \citep{Shadlen2001, Gold2007, Latimer2015}. But other sequential sampling models cannot be used to study magnitude estimation, and the AQ model bridges that gap. It does so by separating the objective properties of the environment, on the one hand, from the subjective properties the cognitive system perceives, on the other.

\subsection{The Environment}
\noindent The AQ assumes the environment, either the present environment or one in the future, contains $C$ discrete structures that signal information. These structures are elements of a system and the information they signal relates to an objective property, $D$, of that system. 

Across elements, signals may conflict. For example, an ox (the system) can have fully developed horns (one element), but also low height (another element). The first element signals greater weight (the objective property of the system), while low height indicates the opposite.

Elements signal $D$ perfectly when summed correctly, $\sum _{j=1}^CC_j$, but they also signal $D$ when the deviation from their corresponding mean element is summed,

\begin{equation}
\label{eq:AQenv}
D -\bar{D} = \sum _{j=1}^C(C_j- \bar{C}_j),
\end{equation}

\noindent where $\bar{C}_j$ is the mean value of the $j'th$ element of the system, and $\bar{D}$ is the mean value of the objective property across the category of system in question. For example, all the elements of an ox - its shoulders, its rump, its legs, and so on - together signal what it weighs without any discrepancy. However, so does the typical weight of oxen, together with all the differences between each element and the mean across their corresponding elements in the category "ox". This assumed feature of the environment is crucial for how the cognitive system is thought to form estimates of magnitude, given its limitations.  

Finally, and defining for the AQ, the environment is simplified by the assumption $C_j- \bar{C}_j = \pm v$, where $v$ is a constant for the amount of information that each element signals about $D$

\subsection{The Cognitive System}
\noindent The cognitive system's imperfections affect how it detects signals from the environment. The imperfections not only force the cognitive system to attend elements sequentially but also introduce noise to its signal detection process. That noise is the conjectured source of uncertainty in the estimates produced by the cognitive system, and noisy signal detection is therefore the basic reason why the KF should work on judgments of magnitude.

Adaptation Level Theory \citep{Helson1947} and Norm-Based \citep{Rhodes2005} coding together explain the details. Due to its imperfections, the cognitive system cannot process all the signals at once. However, by presuming the objective property equals the norm of prior experience, the cognitive system copes.

To presume the objective property equals the norm of prior experience works because the objective property often is near that level when the environment has regularities. Rather than treating every new situation as new, the cognitive system instead focuses its limited resources on detecting how elements deviate from their norms and thereby works out how it should adjust away from the overall norm (i.e., the norm at the system level) given the evidence. Even when detection is coarse, this strategy works well in regular environments when detection errors are unsystematic, which the AQ assumes they are.

Subjects often form adaptation levels during psychophysical tasks \citep{Helson1947, Berniker2010}, and scans of neurons during these tasks offer evidence of norm-based coding in the brain \citep{Leopold2006, Loffler2005}. Two pools of competing neurons that have the same level of activity at the adaptation level are involved. One of these pools responds with increasing average intensity to magnitudes greater than level, while the other decreases its average intensity, thus forming an X-shaped pattern. Through their activity, these pools thus offer evidence about the deviation between the presented stimuli, and the norm. 

The AQ combines adaptive level theory, norm-based coding, and sequential sampling to model magnitude estimation as an outcome of noisy evidence accumulation by neurons. Neurons introduce noise for two related reasons. First, when they have the highest response, they win the right to define how elements deviate from their norm, and second, neurons respond to stimulus with variation and may sometimes win without basis.

More specifically, the cognitive system attends each $C_j$ sequentially to estimate $D$, comparing each $C_j$ to its corresponding mean, which the cognitive system is assumed to have learned. For each $C_j$ sampled, response variance among the competing pools of neurons creates the possibility that neurons that support the idea $C_j < \bar{C}_j$ will respond with least intensity, although $C_j$ < $C_j$, and neurons supporting the idea $C_j > \bar{C}_j$ will respond with least intensity, although $C_j > \bar{C}_j$. Since the pools of neurons are assumed to win by having the highest response, both cases will result in the wrong detection of $C_j$ and will cause an increase in the weight of evidence by $+v$ when the proper incremental change was $-v$ and vice versa.

The final but most defining feature of the AQ is how simply it captures the brain's ability to detect signals. The AQ assumes the cognitive system detects any element correctly with probability $p$, and any element wrongly with probability $1-p$. When $p=1$, the accumulation of evidence leads to an estimate equal to $D$ because the evidence gathered by the pools of neurons will correspond to the signals sent by the environment. For $p<1$, however, the cognitive system can produce many different estimates of $D$, thus causing an uncertain link between the cognitive system's estimate of $D$, and what magnitude $D$ has.

The AQ becomes a Quincunx because of this final assumption about signal detection. However, as its name suggests, the AQ is an augmented version of Galton's original device, which he built in 1873 to demonstrate the Central Limit Theorem \citep{Galton1894}. In the AQ, "balls" move around displaced "pins" and indicate "success" by heading left at some junctures, and right at others, whereas Galton assigned "success" to one direction only, and this small difference is decisive for applying the device to model human cognition.

As shown in detail elsewhere \citep{Nash2017}, the AQ predicts that estimates by the cognitive system of how unusual something is can be described by a random variable $e_d$ with the following mean, variance, skew, and kurtosis

\begin{eqnarray}
\label{eq:mean}
\mu &=& (2 p -1) t v \\
\label{eq:variance}
\sigma^2 &=& 4 C (1-p) p v^2 \\
\label{eq:skewness}
\gamma &=& -\frac{2\mu}{C \sigma} \\
\label{eq:kurtosis}
\kappa &=& 3+\frac{4 v^2}{\sigma ^2}-\frac{6}{C}
\end{eqnarray}

From here, the cognitive system estimates $D$ without much effort by adding $\bar{D}$ to $e_d$. Effectively, $\bar{D} + e_d$ offers researchers an exciting set of grounded predictions about the uncertainty of estimates by the brain. As such, we can use the AQ not only to implement the KF and remove some of these uncertainties but also predict and understand the scope and limitations of the KF when estimates come from humans.

\section{Predicting the Wisdom of a Kalman Crowd using the AQ}
\noindent Let us, therefore, return to the market and the problem faced by the butcher, the farmer, and the official, who must optimally combine their judgments about how much the ox weighs.

As before, we start by looking at the initial problem faced by the butcher and the farmer, but now restate equations \ref{eq:4} and \ref{eq:5} using the AQ. Doing that yields

\footnotesize
\begin{equation}
\label{eq:9}
w_1^* = \frac{4 C \left(1-p_2\right) p_2 v^2+\left(t v-\left(2 p_2-1\right) t v\right) \left(\left(2 p_1-1\right) t v-\left(2 p_2-1\right) t v\right)}{4 C \left(1-p_1\right) p_1 v^2+4 C \left(1-p_2\right) p_2 v^2+\left(\left(2 p_1-1\right) t v-\left(2 p_2-1\right) t v\right){}^2}
\end{equation}
\normalsize

\noindent and

\footnotesize
\begin{equation}
\label{eq:10}
w_2^* = 1 - \frac{4 C \left(1-p_2\right) p_2 v^2+\left(t v-\left(2 p_2-1\right) t v\right) \left(\left(2 p_1-1\right) t v-\left(2 p_2-1\right) t v\right)}{4 C \left(1-p_1\right) p_1 v^2+4 C \left(1-p_2\right) p_2 v^2+\left(\left(2 p_1-1\right) t v-\left(2 p_2-1\right) t v\right){}^2},
\end{equation}
\normalsize

\noindent which provides insights beyond what equations \ref{eq:4} and \ref{eq:5} offer. Where those equations tell us that we should place the most weight on the estimate that is the most certain, holding the levels of bias constant, equations \ref{eq:9} and \ref{eq:10} go deeper. They hold the more profound message that we should place the most weight on estimates produced through more reliable signal detection because that process is not only the most certain but also the least biased except when the objective property is typical. Indeed, regarding bias, equations \ref{eq:9} and \ref{eq:10} give us more than the general advice that we should place the least weight on the most biased estimate, holding uncertainty constant. Instead, we now learn that since greater bias occurs when the objective property is increasingly extreme, and since estimates produced through more reliable signal detection are less biased in all those situations, then we should place even more weight on such estimates when the objective property is more unusual. Simply put, adept judges should be headed more when things are unusual.

Unfortunately, following that last piece of advice is practically impossible because calibrating the weight placed on an estimate based on how unusual the objective property is, implies knowing that magnitude already, which is what we are trying to find out. Consequently, we must assume the objective property is typical, which means setting $t=0$. However, by making that assumption, we are instantly assuming all estimates are unbiased, which means the predicted nature of magnitude estimation has forced us to introduce to the KF what the AQ predicts will be an imprecision. Nevertheless, we set $t=0$ and thus obtain an exact correspondence to the Kalman Gain equations \ref{eq:6} and \ref{eq:7} applicable to unbiased electronic sensors. More precisely, equations \ref{eq:9} and \ref{eq:10} simplify noticeably to

\begin{equation}
\label{eq:11}
w_1^* = \frac{\left(1-p_2\right) p_2}{\left(1-p_1\right) p_1+\left(1-p_2\right) p_2}
\end{equation}

\noindent and

\begin{equation}
\label{eq:12}
w_2^* = 1 - \frac{\left(1-p_2\right) p_2}{\left(1-p_1\right) p_1+\left(1-p_2\right) p_2}.
\end{equation}

At this point we notice the number of elements in the system, $C$, and the evidence carried by each element, $v$, have been eliminated. Accordingly, what the AQ tells us is that when the subject of estimation is typical, nothing particular about its character is essential for combining estimates optimally; the relative reliability of signal detection, $p_1$ versus $p_2$, is all that matters.

\subsection{Verification by a third party:} 
\noindent How the butcher and the farmer, in our particular case, should know each other's values of $p$ is something else entirely. In the case of proper scales, knowing the uncertainty of estimates is easier because intensive testing increases the number of observation beyond the point where further evidence has little effect. In contrast, because we cannot run such tests on people, the limited number of observed estimates affects our knowledge about how uncertain estimates are. Consequently, our uncertainty about uncertainty reduces the prospect of placing the correct weights. Nevertheless, we will defer this critical issue for a little while yet. Instead, we will continue to assume that the butcher, the farmer, and indeed the official who must verify that everything is in the best order, know the inherent uncertainties of judgments.

Like before, we therefore prepare for the official's update by first substituting the Kalman Gain equations (equations \ref{eq:11} and \ref{eq:12}) into equation \ref{eq:3} while remembering to set $t = 0$. Performing these substitutions gives

\begin{equation}
\label{eq:13}
MSE_{KF} = 4Cv^2\frac{\left(1-p_1\right) p_1 \left(1-p_2\right) p_2}{\left(1-p_1\right) p_1+\left(1-p_2\right) p_2}.
\end{equation}

\noindent where $C$ and $v$ provide some useful insight. Simply put, some things are small while other things are more prominent, and for that reason, the number of elements and the information that each element provides will affect the size of errors that people make.

\subsection{The AQ's Property of Closure:}
\noindent We have arrived at a crucial point in this paper. Earlier we noted the MSE of the combined estimate is equal to the uncertainty of this estimate when it is unbiased. Furthermore, we noted how the assumption of Gaussian distributed estimates offers the possibility of modeling the combined estimate using a distribution that is also Gaussian, thus resetting the process mathematically after every update. However, since the property of closure is rare, and since we must use the AQ for theoretical reasons, producing an increasing amount of unruly math after every step looks more likely.

Thankfully, however, that is wrong. The AQ offers the possibility of treating every combined estimate as though a person with an equivalent level of signal detection ability had produced it. We can do that because when we know the value of $p$, then we also know the mean and variance of the estimate, which are the essential ingredients we need to compute MSE. As such, we can set equation \ref{eq:13} equal to variance equation \ref{eq:variance} and solve for $p$ to get

\begin{equation}
\label{eq:14}
p_{KF} = \frac{1}{2} + \frac{1}{2} \sqrt{1 - 4\frac{\left(1-p_1\right) p_1 \left(1-p_2\right) p_2}{\left(1-p_1\right) p_1+\left(1-p_2\right) p_2}}.
\end{equation}

\noindent We can interpret equations \ref{eq:11} to \ref{eq:14} from two perspectives, the first of which is interesting from an organizational view. If we know the uncertainty of estimates by individuals in a large group, then we can arrange this group into pairs, and work out how much weight we should place on estimates by members of each pair in order to minimize the MSE of the pair's combined estimates. We can then state the combined ability of each pair as if they were individuals, and in doing so, can virtually halve the size of the group but with new aggregate units that all generate less error than their constituents. At that point, we can continue the process as though starting over mathematically, and form pairs using the aggregate units, and so on until we have assembled all people into one entity that generates estimates with the least possible MSE overall. Moreover, no matter how we form our pairs, we would reach the same minimum. Alternatively, we may choose to interpret the function of equations \ref{eq:11} to \ref{eq:14} in the standard way, as working recursively through estimates by members of the group.

Of course, the official who wants to obtain the best possible estimate of an ox by combining his judgment with that made by a butcher and a farmer, or the manager who wishes to obtain the best possible estimate from hundreds of employees under her management, would want to know how the uncertainty about uncertainty affects the KF. The time has come to look at that crucial issue.

\subsection{The $KF_u$ vs. the $KF_c$:} 
\noindent Let us suppose a manager wants to use the KF on estimates by her employees about the earnings per share of Coca-Cola. How should the manager proceed when there is no information about how the employees' estimates, and Coca-Cola's earnings per share, relate?

It seems evident that what the manager should do is learn the relationship, which essentially means observing how the estimates of different employees relate to earnings across many observations. As such, we can reframe the problem to one of learning how the KF can be expected to perform as the observed number estimates increases, and when the manager uses the mean of the observed squared errors to set $w_1$ and $w_2$ with the least margin to $w_1^*$ and $w_2^*$.

For three reasons, we are going to keep the assumption that estimates are unbiased. First, that links the problem to what we assumed above. Second, it makes the mean of the observed squared errors equivalent to a sample of variance, and third, it offers the opportunity to use Cochran's Theorem, which relates to the distribution of such samples when the random variable (i.e., the estimate of magnitude in the present case) follows a Gaussian distribution. More precisely, when samples of $n$ observations are taken from the Gaussian distribution with variance $\sigma^2$ then Cochran's Theorem states that $\frac{n S^2}{\sigma ^2}\sim \chi _{n-1}^2$, where $S^2$ is the sample variance. Consequently

\begin{equation}
\label{eq:15}
S^2 \sim \Gamma \left(\frac{n-1}{2},\frac{2 \sigma ^2}{n}\right).
\end{equation}

At this point, the reader might ask how we can apply Cochran's Theorem in the current situation. After all, we took deliberate steps to model magnitude estimates as random variables that capture aspects of both the environment and the cognitive system, instead of merely assuming they follow the Gaussian distribution. The answer is quite simply that $e_d$ tends towards the Gaussian distribution under particular circumstances, which allows us to proceed as stated while harvesting the deeper insight, which the AQ provides. More precisely, if we assume the environment has an infinite number of elements and then assume each of these elements tend to carry an infinitesimally small amount of evidence, then the moment generating function of $e_d$ becomes

\begin{equation}
\label{eq:16}
M_{x} = e^{-2 (p-1) p x^2},
\end{equation}

which equals the moment generating function of the Gaussian distribution with zero mean, and with variance given by \ref{eq:variance}. In other words, $e_d$ is Gaussian under the specified assumptions. We can, therefore, make these assumptions about $C$ and $t$ and proceed to compare the MSE produced by the KF when weights on pairs of estimates are placed certainly versus uncertainly. Denoting the MSE of the uncertain KF by $KF_{u}$, and the MSE of the certain KF by $KF_{c}$, the difference in MSE generated by these two filters is given by

\tiny
\begin{equation}
\label{eq:17}
KF_{u} - KF_{c}  = \left(\sigma _1^2 \left(\frac{S^2_2}{S^2_1+S^2_2}\right){}^2+\sigma _2^2 \left(1-\frac{S^2_2}{S_1+S^2_2}\right){}^2\right)-\left(\sigma _1^2 \left(\frac{\sigma _2^2}{\sigma _1^2+\sigma _2^2}\right){}^2+\sigma _2^2 \left(1-\frac{\sigma _2^2}{\sigma _1^2+\sigma _2^2}\right){}^2\right)
\end{equation}
\normalsize

\noindent which is a random variable due to \ref{eq:15}. Setting $n=2$, because this corresponds to the level in our upcoming empirical study, we find the expectation of equation \ref{eq:17} to be

\tiny
\begin{equation}
\label{eq:18}
E[KF_{u} - KF_{c}]  = \frac{\sigma _1^2 \left(-5 \left(\sigma _2^2\right){}^{5/2} \left(\sigma _1^2\right){}^{3/2}+8 \sigma _1^2 \sigma _2^6+\sqrt{\frac{\sigma _2^{18}}{\sigma _1^2}}+\sqrt{\sigma _1^{10} \sigma _2^6}-5 \sqrt{\sigma _1^2 \sigma _2^{14}}\right)}{2 \sigma _2^2 \left(\sigma _1^2-\sigma _2^2\right){}^2 \left(\sigma _1^2+\sigma _2^2\right)}.
\end{equation}
\normalsize

Finally, making use of the fact that variance equation \ref{eq:variance} simplifies to $4(1-p)p$ when $C = \infty$ and $v\to 0$, we substitute $4(1-p_1)p_1$ for $\sigma _1^2$ and $4(1-p_2)p_2$ for $\sigma _2^2$ in \ref{eq:18} to generate the top panel of Figure 1.

\begin{figure*}
\begin{center}
\caption{\em The Certain Kalman Filter vs. Other Rules}
\centerline{\includegraphics[width=0.65\textwidth]{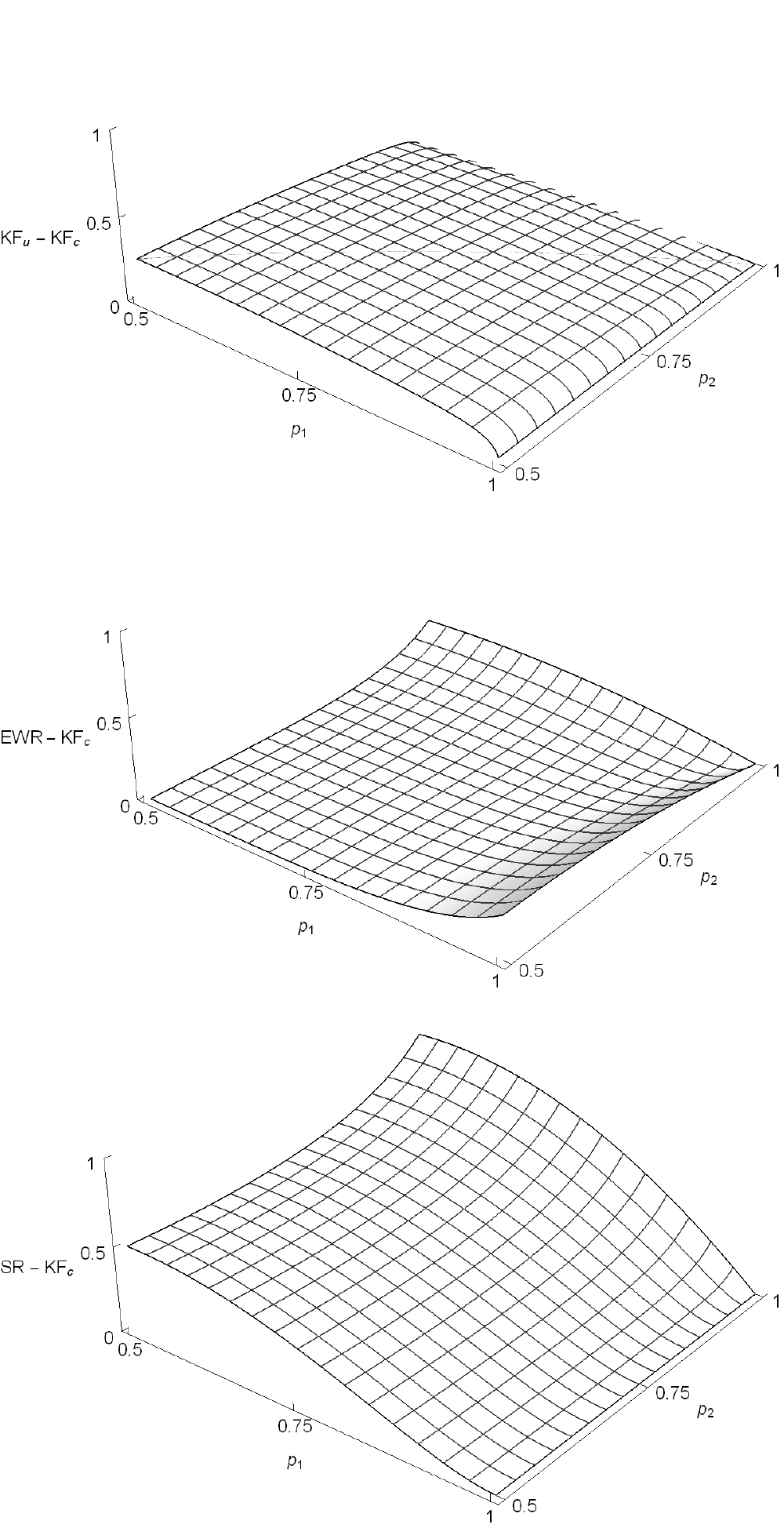}}
\label{fig:1}
\end{center}
\end{figure*}

We learn two things from Figure 1, and both say something about how well we can expect the KF to work in practice. First, the certain KF beats the uncertain KF mainly when unreliable signal detection defines the sources of estimates. Second, the certain KF beats the uncertain KF by less when good signal detection defines one of the sources, even when the reliability of the other is constant. In both cases, signal detection determines the inherent uncertainty of estimates, which in turn determines the expected difference between the inherent and inferred uncertainty. More precisely, the expected difference can be shown to be $\frac{4(1-p)p}{n}$. Accordingly, we state the following predictions:

\bigbreak

\begin{hypothesis}
\textbf{Prediction 1:} The KF will function better (worse) when people are adept (novice).
\end{hypothesis}

\bigbreak

\begin{hypothesis}
\textbf{Prediction 2:} The KF will function better (worse) when people have easier (harder) estimation problems.
\end{hypothesis}

\bigbreak

\begin{hypothesis}
\textbf{Prediction 3:} The KF will function better (worse) when people have shorter (longer) forecast horizons.
\end{hypothesis}

\bigbreak

These predictions are the same from the AQ's perspective. They assert the KF functions better (worse) when $p$ is larger (smaller) since the difference between the inherent and inferred uncertainty $\frac{4(1-p)p}{n}$ is smaller (larger) in those situations, which results in combinations of estimates that are closer to (farther from) optimal.

\subsection{The $EWM$ vs. the $KF_u$:} 
\noindent The certain KF beats the EWR for any pair of $p_1$ and $p_2$ (Figure 1, middle), but more interesting is the performance of the uncertain KF versus the EWR,

\tiny
\begin{equation}
\label{eq:19}
EW - KF_{u}  = \left(\left(\frac{1}{2}\right)^2 \sigma _1^2+\left(\frac{1}{2}\right)^2 \sigma _2^2\right)-\left(\sigma _1^2 \left(\frac{S_2}{S_1+S_2}\right){}^2+\sigma _2^2 \left(1-\frac{S_2}{S_1+S_2}\right){}^2\right).
\end{equation}
\normalsize
Like equation \ref{eq:16}, $EW - KF_{u}$ is a random variable due to equation \ref{eq:15}, but we can examine its expected value. Setting $n=2$, that value is

\tiny
\begin{equation}
\label{eq:20}
E[EW - KF_{u}]  = \frac{12 \left(\sigma _2^2\right){}^{5/2} \left(\sigma _1^2\right){}^{3/2}-2 \left(\sigma _2^2\right){}^{3/2} \left(\sigma _1^2\right){}^{5/2}+\sigma _2^8-5 \sigma _1^2 \sigma _2^6-5 \sigma _1^4 \sigma _2^4+\sigma _1^6 \sigma _2^2-2 \sqrt{\sigma _1^2 \sigma _2^{14}}}{4 \sigma _2^2 \left(\sigma _1^2-\sigma _2^2\right){}^2}.
\end{equation}
\normalsize

Finally, by substituting $4(1-p_1)p_1$ for $\sigma _1^2$ and $4(1-p_2)p_2$ for $\sigma _2^2$ into equation \ref{eq:20}, we can generate Figure 2.

\begin{figure*}
\begin{center}
\caption{\em The Equal-Weight Model vs. the Uncertain Kalman Filter}
\centerline{\includegraphics[width=1\textwidth]{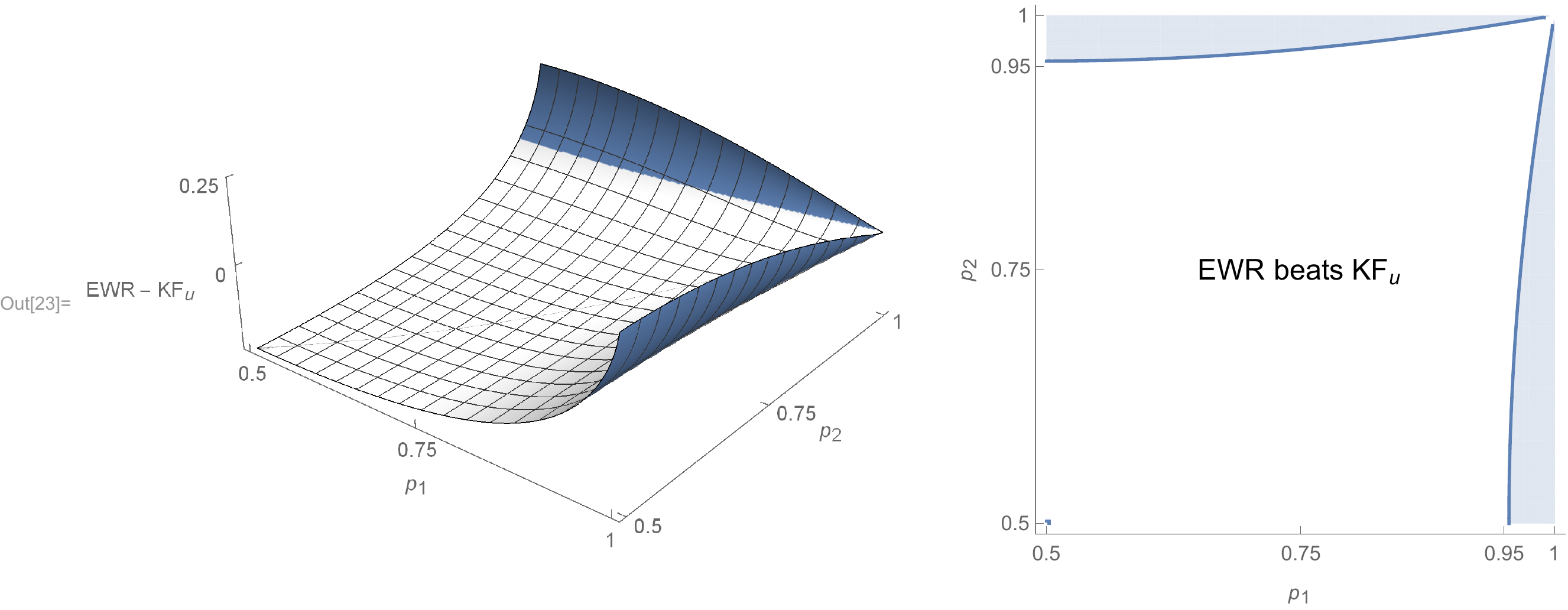}}
\label{fig:2}
\end{center}
\end{figure*}

From the right side of Figure 2, we notice that at the studied level of uncertainty, the EWR beats the uncertain KF for most $p_1$ and $p_2$ combinations. However, we also notice the uncertain KF beats the EWR when highly reliable signal detection processes define one of the estimates to be combined. Again, this makes good sense because smaller inherent uncertainty defines such estimates. With smaller inherent uncertainty, an observer will experience less uncertainty about uncertainty, than when this basic uncertainty is more substantial. The weight placed on the certain estimate will thereby tend to be correct. However, so will the weight placed on the other, even though this estimate is more uncertain, because the weights sum to one. Accordingly, the KF gets this type of situation right, whereas the EWM gets it wrong per definition. 

The left side of Figure 3 offers further insight. Not surprisingly, the EWR outperforms by most when $p_1 = p_2$ because here the rule is optimal, while the uncertain KF only hits optima on average. Perhaps less obvious, however, is the observation that although the EWR beats the uncertain KF for most $p_1$ and $p_2$ combinations, the effect is only marginal when that happens. In contrast, in those regions where the uncertain KF beats the EWR, it does so with considerable margin. Accordingly, we state the following predictions: 

\bigbreak

\begin{hypothesis}
\textbf{Prediction 4:} The KF will lose by a small margin to the EWR when people have similar levels of adeptness.
\end{hypothesis}

\bigbreak

\begin{hypothesis}
\textbf{Prediction 5:} The KF will lose by a small margin to the EWR when people are diverse but novice on average.
\end{hypothesis}

\bigbreak

\begin{hypothesis}
\textbf{Prediction 6:} The KF will beat the EWR by a large margin when people are diverse but adept on average.
\end{hypothesis}

\bigbreak

\begin{hypothesis}
\textbf{Prediction 7:} The KF will beat the EWR by a large margin when some people are adept, and others are novice.
\end{hypothesis}

\bigbreak

Predictions 5 and 6 can be detailed further in terms of problem difficulty, but we leave that implied by predictions 1 - 3. 

\subsection{The $SR$ vs. the $KF_u$:} 
\noindent When we know the uncertainty of estimates, then combining the estimates available using the KF is better than waiving this opportunity (Figure 1, bottom). However, beyond some difference between the inherent and inferred uncertainty, the weight we give to pairs of estimates may be so far from optimal that some estimates are better left as they are. These include individual estimates by adept people, but they also include estimates by many people that we have already combined using some rule, including the CWM or the KF itself. In both cases, we are dealing with the inclusion of some subset of estimates, and the rejection of others. We therefore call this rule the Subset Rule (SR).

To see how the uncertain KF performs against the SR, we will use $p_1$ to capture the MSE of the subset estimate. We will then compare that MSE to the one we could obtain if we instead fused the subset estimate with another estimate (defined by $p_2$) using the uncertain KF. As before, however, our first step is to express the difference in MSE most generally. That difference is the random variable

\begin{equation}
\label{eq:21}
SR - KF_{u}  = \sigma _1^2-\left(\sigma _1^2 \left(\frac{S_2^2}{S_1^2+S_2^2}\right){}^2+\sigma _2^2 \left(1-\frac{S_2^2}{S_1^2+S_2^2}\right){}^2\right),
\end{equation}

which has an expected value of

\begin{equation}
\label{eq:22}
E[SR - KF_{u}]  = \frac{-\sigma _2^4+3 \sigma _1^2 \sigma _2^2+2 \sqrt{\sigma _1^6 \sigma _2^2}-2 \sqrt{\sigma _1^2 \sigma _2^6}}{2 \sqrt{\frac{\sigma _2^2}{\sigma _1^2}} \left(\sigma _1^2+\sigma _2^2+2 \sqrt{\sigma _1^2 \sigma _2^2}\right)},
\end{equation}
\noindent when $n=2$. Finally, we substitute $4(1-p_1)p_1$ for $\sigma _1^2$ and $4(1-p_2)p_2$ for $\sigma _2^2$ into equation \ref{eq:22} to generate Figure 3.

\begin{figure*}
\begin{center}
\caption{\em The Subset Rule vs. the Uncertain Kalman Filter}
\centerline{\includegraphics[width=1\textwidth]{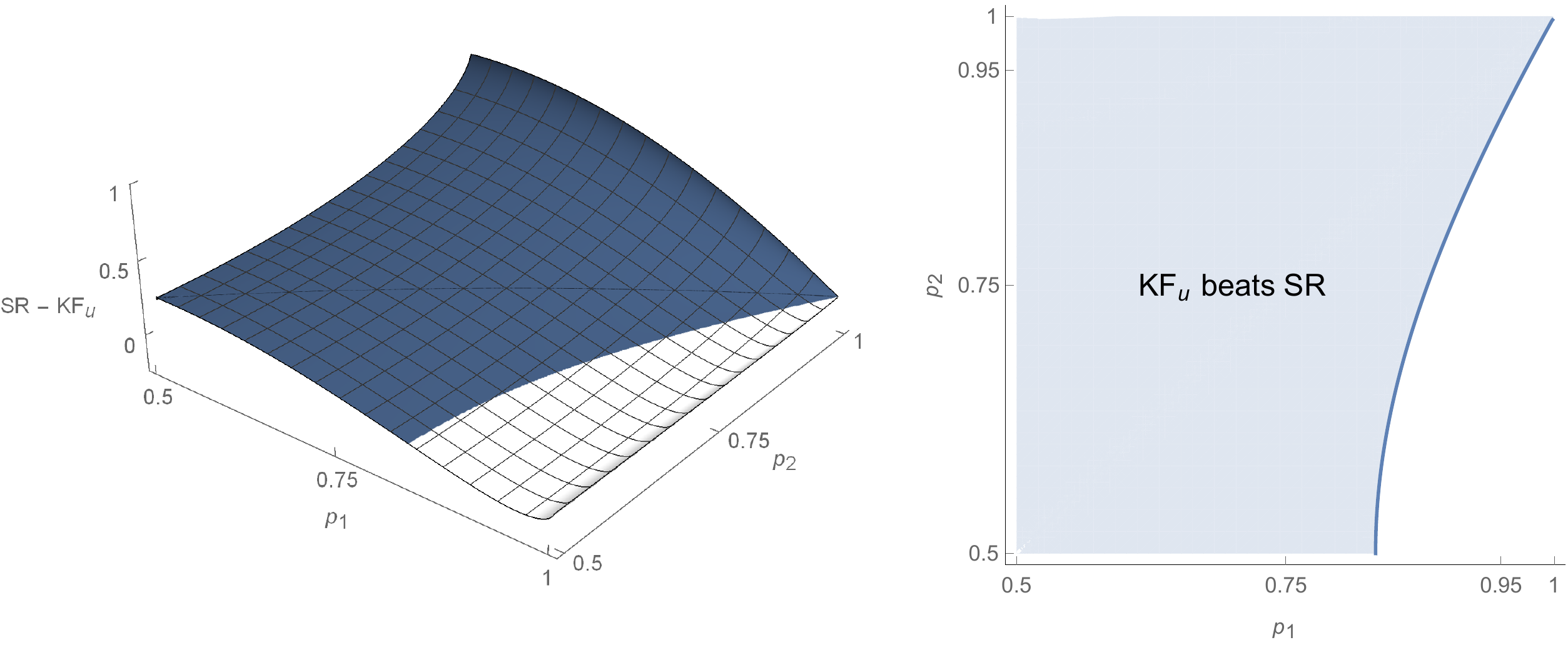}}
\label{fig:3}
\end{center}
\end{figure*}

As anticipated from the bottom of Figure 1, the right side of Figure 3 reveals that waiving the opportunity to combine the available estimates requires that a high level of $p_1$ defines the subset. However, even when that condition is met, waiving the opportunity to combine the available estimates requires a second condition. It also requires that the other estimate is defined by a small to a moderate level of $p_2$. If not, then the uncertain KF will beat the SR. Moreover, as the left side of Figure 3 shows, when $p_2$ is high and becomes exactly so high that the uncertain KF just becomes superior, then this marginal effect is quite substantial, at least compared with the similar situation where $p_2$ is small or moderate. Accordingly, we state the following predictions:

\bigbreak

\begin{hypothesis}
\textbf{Prediction 8:} The KF will beat the SR unless the subset's estimate is defined by high reliability, and the excluded estimate is not.
\end{hypothesis}

\bigbreak

\begin{hypothesis}
\textbf{Prediction 9:} When the KF beats the SR, it will do so with greater effect when the excluded estimate is defined by high reliability.
\end{hypothesis}

\section{Study: Professional Forecasters of Economic Variables}
\noindent We used data from the Survey of Professional Forecasters (SPF) to see how well the KF works on estimates of magnitude made by humans. The American Statistical Association and the National Bureau of Economic Research started the survey in 1968, but the Federal Reserve Bank of Philadelphia (FRBP) took over in 1990. SPF is the oldest quarterly survey of macroeconomic forecasts in the United States.

The SPF offers an excellent opportunity to examine how well the KF works in a setting not too different from the ones that managers face. The survey concerns economic variables that affect the environment of firms, and its participants have strong knowledge of finance, economics, and banking. As such, the forecasters in the survey are like many managers. Moreover, the data from the survey is plentiful, but also sparse due to turnover among its participants. Since turnover among managers also characterizes firms, this feature adds unique realism — it changes the knowledge base of the survey through time and creates practical obstacles to harnessing the wisdom of crowds that firms also have. Furthermore, the survey involves forecasting over different horizons and thereby has various degrees of difficulty. Finally, because the survey is about the US economy, it mirrors the dynamic forecasting problems that managers solve through the financial year.

Besides being very suitable to the management domain, the SPF also has features that make it very relevant for studying different aggregation rules, including the KF. Most notably, participants are anonymous, which reduces the incentive to conform and makes their estimates more independent. Moreover, given their adeptness, we can expect them to be less biased than novices.

Last, but not least, since forecasts are known \citep{Stark2010} to involve the use of intuitive judgment, using the AQ to inform our study makes good sense.

\subsection{Data Collection}
\noindent We analyzed quarterly forecasts of 12 variables relating to the US economy across five horizons, these horizons being 1 to 5 quarters (steps) ahead. The selected economic variables were chosen because the FRBP offers vintage data for them. Vintage data is crucial because it shows the historical data available to forecasters as they forecast, as well as the actual value of the economic variable as the FRBP first reports them, absent any official changes that forecasters could not foresee. To create consistency across variables, we converted all forecasts and realized values to yearly percentage changes, except for civilian unemployment, which the FRBP already states in percentage terms. Since the start in 1968, the average number of forecasters in each survey has ranged from 30 to 36, whose typical tenure has varied between 11 to 18 surveys. The turnover rate of forecasters has varied 15 to 23 percent. In total, the data consists of 326238 forecasts across 9715 surveys. Table 1 and Table 2 offer further details of the surveys and some descriptive statistics.

\begin{table}
\footnotesize
\centering
\caption{The Twelve Economic Variables and their Abbreviations}
\begin{tabular}{ll} 
\hline
\textbf{Variable}                                & \textbf{Abbreviation}  \\ 
\hline
Nonfarm Payroll Employment                       & EMP                    \\
Housing Starts                                   & HOUSING                \\
Nominal Gross National Product                   & NOUTPUT                \\
Price Index for Gross National Product           & PGDP                   \\
Industrial Production Index                      & INDPROD                \\
Real Personal Consumption Expenditures           & RCONSUM                \\
Real Federal Government Consumption Expenditures & RFEDGOV                \\
Real Gross National Product                      & RGDP                   \\
Real Nonresidential Fixed Investment             & RNRESIN                \\
Real Residential Fixed Investment                & RRESINV                \\
Real State and Local Government Consumption      & RSLGOV                 \\
Civilian Unemployment Rate                       & UNEMP                  \\
\hline
\end{tabular}
\end{table}

\begin{table}
\footnotesize
\centering
\caption{Descriptive Statistics for the Survey of Professional Forecasters}

\begin{center}
    \addtolength{\leftskip} {-2cm} 
    \addtolength{\rightskip}{-2cm}
\begin{tabular}{lllll} 
\hline
\textbf{Abbreviation}     & \textbf{No. surveys} & \textbf{Median No. judges} & \textbf{Median tenure} & \textbf{Turnover rate}  \\ 
\hline
EMP                       & 275                  & 36                         & 18                     & 0.15                    \\
HOUSING                   & 965                  & 34                         & 15                     & 0.23                    \\
NOUTPUT                   & 975                  & 35                         & 16                     & 0.22                    \\
PGDP                      & 975                  & 35                         & 15                     & 0.22                    \\
INDPROD                   & 975                  & 33                         & 15                     & 0.23                    \\
RCONSUM                   & 720                  & 33                         & 11                     & 0.19                    \\
RFEDGOV                   & 720                  & 31                         & 12                     & 0.19                    \\
RGDP                      & 975                  & 36                         & 15                     & 0.22                    \\
RNRESIN                   & 720                  & 31                         & 11                     & 0.19                    \\
RRESINV                   & 720                  & 32                         & 11                     & 0.19                    \\
RSLGOV                    & 720                  & 30                         & 12                     & 0.19                    \\
UNEMP                     & 975                  & 36                         & 16                     & 0.22                    \\ 
\hline
                          &                      &                            &                        &                         \\
\textbf{Total surveys:}   & 9715                 &                            &                        &                         \\
\textbf{Total forecasts:} & 326238               &                            &                        &                        
\end{tabular}
\end{center}
\end{table}

\subsection{Method}
\noindent We treated the data like a company would as it receives forecasts from different employees about business variables across time. Accordingly, our "company" could only update its measures of individual performance using the data available before the latest survey. We assumed the company updated two such measures for each economic variable and forecast horizon and then used these to weigh the estimates that "employees" (i.e., the professional forecasters) provided. All estimates by forecasters with fewer than two observations were discounted completely by all aggregation rules. That rule is identical to the one used by \cite{Budescu2015} in their study of survey data from the European Central Bank.   

One of the performance measures used was $p$ from the AQ model, which was estimated using the latest mean of the forecaster's squared errors. More precisely, because we had to assume that estimates were unbiased for theoretical reasons\footnote{As stated earlier, the AQ predicts that estimates are biased when the objective property is unusual. However, we cannot grant forecasters any knowledge about the objective property and must, therefore, assume the objective property equals the norm. In turn, that assumption leads to the assumption of unbiased estimates, since the AQ makes that prediction when the objective property has that magnitude}, the MSE and variance equation \ref{eq:variance} become equal, such that 

\begin{equation}
\label{eq:23}
MSE = 4 C (1-p) p v^2\Rightarrow p = \frac{1}{2} + \frac{\sqrt{C v^2 \left(C v^2 - MSE\right)}}{2 C v^2}.
\end{equation}

Accordingly, as new squared errors by the particular forecaster became available, the estimate of the forecaster's $p$ could be updated and used to determine what weight the estimate from the forecaster should receive in the KF on the next survey where the forecaster was active (see equations \ref{eq:11} to \ref{eq:14} for detailed instructions). However, as equation \ref{eq:23} reveals,  the constants $C$ and $v$ had to be set first. What these are is arbitrary for the KF since they apply to all forecasters across all periods (see equation \ref{eq:14} for details). However, we set $C = 1$ in all surveys and across all horizons to provide the opportunity of direct comparison. Moreover, we used the same rule across all surveys to set $v$. Specifically, $v$ was set by first assuming the norm of each economic variable equaled their arithmetic average in the vintage data. Then, we identified the most extreme observation for each time-series. Finally, we computed the absolute difference of this extreme value from the norm and set $v$ to that value\footnote{Although not crucial for the KF, the way $v$ and $C$ are decided is consistent with the workings of the Quincunx. In essence, what we did was set the value of $C$ and $v$ such that balls falling through the device could reach every one of its compartments. Had we instead assumed that $v$ was smaller, then the most extreme observations could, in strict theoretical accordance with the AQ, not be inferred because the $C = 1$ element would not carry enough information}. 

The second performance measure used was contribution (CW), which forms the basis of the CWM. The contribution of each forecaster just before the current survey is the average difference between the MSE of the EWR with and without the forecaster included, across all surveys where the forecaster has been active. Formally, that is expressed as

\begin{equation}
\label{eq:24}
CW_j = \frac{1}{N_j} \sum _{i=1}^{N_j} \left(EWR_i-EWR_i^{-j}\right)
\end{equation}

where $j = 1,...,J$ denotes the particular forecaster, and $N_j$ denotes the surveys where the forecaster was active. CW formed the basis of the CWM in the current survey in three steps. First, the subset of active forecasters with positive CW were identified. Second, the corresponding positive $CW_j$'s were normalized, and finally, these normalized $CW_j$'s were used to assign a weight to the predictions of the associated forecaster. As such, the CWM is a weighted average computed at once, whereas the KF is computed recursively. 

\subsection{Comparison of the Aggregation Rules}
\noindent Table 3 compares the performance of four aggregation rules. These rules are the EWM, the KF, the SR represented by the CWM, and the KF+. The KF+ is like the CWM but with an important difference. It works on the same subset of forecasts, but where the CWM weighs forecasts using the normalized positive contribution measures, the KF+ assigns weights using the KF based on the estimated $p$'s. Consequently, we can attribute any difference in performance between the CWM and the KF+ to how optimally they assign weights.

\begin{table}
\footnotesize
\centering
\caption{RMSEs, and DM p-values versus CWM, for Different Variables and Forecast Horizons}
\begin{center}
    \addtolength{\leftskip} {-2cm} 
    \addtolength{\rightskip}{-2cm}
\begin{tabular}{llllllllllll} 
\hline
                 &              & \multicolumn{5}{l}{\textbf{RMSE}}                              & \multicolumn{5}{l}{\textbf{Diebold-Mariano p-values}}           \\
                 &              & \textbf{1} & \textbf{2} & \textbf{3} & \textbf{4} & \textbf{5} & \textbf{1} & \textbf{2} & \textbf{3} & \textbf{4} & \textbf{5}  \\ 
\hline
\textbf{EMP}     & \textbf{EWM} & 0.26       & 0.46       & 0.75       & 1.06       & 1.42       & 0.60       & 0.12       & 0.27       & 0.86       & 0.86        \\
                 & \textbf{KF}  & 0.25       & 0.46       & 0.73       & 1.03       & 1.38       & 0.05       & 0.08       & 0.13       & 0.51       & 0.25        \\
                 & \textbf{CWM} & 0.26       & 0.48       & 0.76       & 1.02       & 1.39       &            &            &            &            &             \\
                 & \textbf{KF+} & 0.24       & 0.47       & 0.72       & 0.99       & 1.36       & 0.02 +     & 0.27       & 0.03 +     & 0.05       & 0.08        \\ 
\hline
\textbf{HOUSING} & \textbf{EWM} & 7.27       & 11.32      & 14.86      & 18.71      & 22.73      & 0.59       & 0.90       & 0.97 -     & 1.00 -     & 1.00 -      \\
                 & \textbf{KF}  & 7.66       & 11.14      & 14.47      & 18.13      & 22.01      & 0.83       & 0.72       & 0.81       & 0.95 -     & 0.97 -      \\
                 & \textbf{CWM} & 7.22       & 10.98      & 14.19      & 17.45      & 20.81      &            &            &            &            &             \\
                 & \textbf{KF+} & 7.66       & 10.91      & 14.18      & 17.58      & 20.84      & 0.82       & 0.39       & 0.48       & 0.66       & 0.53        \\ 
\hline
\textbf{NOUTPUT} & \textbf{EWM} & 0.79       & 1.29       & 1.73       & 2.14       & 2.54       & 0.07       & 0.05 +     & 0.91       & 0.94       & 0.98-       \\
                 & \textbf{KF}  & 0.78       & 1.29       & 1.70       & 2.08       & 2.47       & 0.01+      & 0.03 +     & 0.74       & 0.71       & 0.83        \\
                 & \textbf{CWM} & 0.81       & 1.34       & 1.68       & 2.05       & 2.42       &            &            &            &            &             \\
                 & \textbf{KF+} & 0.78       & 1.28       & 1.69       & 2.03       & 2.42       & 0.00 +     & 0.01 +     & 0.61       & 0.33       & 0.47        \\ 
\hline
\textbf{PGDP}    & \textbf{EWM} & 0.40       & 0.67       & 0.94       & 1.27       & 1.62       & 0.00 +     & 0.05 +     & 0.48       & 0.75       & 0.92        \\
                 & \textbf{KF}  & 0.40       & 0.67       & 0.93       & 1.24       & 1.61       & 0.01 +     & 0.03 +     & 0.40       & 0.65       & 0.90        \\
                 & \textbf{CWM} & 0.43       & 0.69       & 0.94       & 1.22       & 1.47       &            &            &            &            &             \\
                 & \textbf{KF+} & 0.41       & 0.67       & 0.91       & 1.18       & 1.52       & 0.03 +     & 0.02 +     & 0.23       & 0.30       & 0.70        \\ 
\hline
\textbf{PROD}    & \textbf{EWM} & 1.28       & 2.25       & 3.21       & 4.08       & 4.86       & 0.93       & 0.97 -     & 0.93       & 0.94       & 0.47        \\
                 & \textbf{KF}  & 1.27       & 2.23       & 3.19       & 4.04       & 4.83       & 0.90       & 0.93       & 0.90       & 0.88       & 0.35        \\
                 & \textbf{CWM} & 1.24       & 2.17       & 3.13       & 3.96       & 4.86       &            &            &            &            &             \\
                 & \textbf{KF+} & 1.25       & 2.19       & 3.13       & 3.97       & 4.75       & 0.78       & 0.79       & 0.54       & 0.48       & 0.05 +      \\ 
\hline
\textbf{RCONSUM} & \textbf{EWM} & 0.69       & 0.96       & 1.20       & 1.50       & 1.84       & 0.12       & 0.62       & 0.12       & 0.11       & 0.16        \\
                 & \textbf{KF}  & 0.69       & 0.96       & 1.18       & 1.49       & 1.94       & 0.13       & 0.64       & 0.08       & 0.09       & 0.47        \\
                 & \textbf{CWM} & 0.80       & 0.95       & 1.30       & 1.62       & 1.95       &            &            &            &            &             \\
                 & \textbf{KF+} & 0.68       & 0.97       & 1.17       & 1.45       & 1.88       & 0.13       & 0.80       & 0.04 +     & 0.03 +     & 0.32        \\ 
\hline
\textbf{RFEDGOV} & \textbf{EWM} & 2.15       & 2.52       & 2.74       & 3.09       & 3.56       & 0.12       & 0.18       & 0.29       & 0.27       & 0.59        \\
                 & \textbf{KF}  & 2.13       & 2.45       & 2.68       & 3.03       & 3.52       & 0.10       & 0.04 +     & 0.13       & 0.16       & 0.49        \\
                 & \textbf{CWM} & 2.28       & 2.60       & 2.80       & 3.19       & 3.52       &            &            &            &            &             \\
                 & \textbf{KF+} & 2.13       & 2.39       & 2.64       & 3.00       & 3.43       & 0.06       & 0.01 +     & 0.06       & 0.05       & 0.28        \\ 
\hline
\textbf{RGDP}    & \textbf{EWM} & 0.72       & 1.24       & 1.67       & 2.09       & 2.47       & 0.07       & 0.01 +     & 0.47       & 0.64       & 0.19        \\
                 & \textbf{KF}  & 0.71       & 1.21       & 1.64       & 2.07       & 2.47       & 0.04 +     & 0.00 +     & 0.22       & 0.53       & 0.21        \\
                 & \textbf{CWM} & 0.75       & 1.37       & 1.68       & 2.07       & 2.56       &            &            &            &            &             \\
                 & \textbf{KF+} & 0.71       & 1.20       & 1.60       & 2.05       & 2.44       & 0.01 +     & 0.01 +     & 0.03 +     & 0.35       & 0.13        \\ 
\hline
\textbf{RNRESIN} & \textbf{EWM} & 2.49       & 3.80       & 5.21       & 6.74       & 8.30       & 0.09       & 0.89       & 0.98 -     & 0.98 -     & 0.97 -      \\
                 & \textbf{KF}  & 2.50       & 3.78       & 5.16       & 6.73       & 8.12       & 0.15       & 0.88       & 0.97 -     & 0.91       & 0.74        \\
                 & \textbf{CWM} & 2.54       & 3.71       & 4.98       & 6.51       & 8.03       &            &            &            &            &             \\
                 & \textbf{KF+} & 2.51       & 3.75       & 5.11       & 6.64       & 7.93       & 0.21       & 0.71       & 0.94       & 0.74       & 0.19        \\ 
\hline
\textbf{RRESINV} & \textbf{EWM} & 2.82       & 4.52       & 6.48       & 8.56       & 10.63      & 0.19       & 0.57       & 0.97 -     & 1.00 -     & 1.00 -       \\
                 & \textbf{KF}  & 2.91       & 4.38       & 6.26       & 8.41       & 10.58      & 0.39       & 0.27       & 0.69       & 1.00 -     & 1.00 -       \\
                 & \textbf{CWM} & 2.94       & 4.49       & 6.19       & 8.06       & 9.81       &            &            &            &            &             \\
                 & \textbf{KF+} & 2.97       & 4.25       & 6.10       & 8.21       & 10.19      & 0.67       & 0.08       & 0.27       & 0.87       & 0.98 -       \\ 
\hline
\textbf{RSLGOV}  & \textbf{EWM} & 0.93       & 1.25       & 1.43       & 1.68       & 2.03       & 0.08       & 0.68       & 0.57       & 0.66       & 0.97 -       \\
                 & \textbf{KF}  & 0.91       & 1.24       & 1.41       & 1.63       & 1.98       & 0.05       & 0.60       & 0.38       & 0.26       & 0.87        \\
                 & \textbf{CWM} & 1.03       & 1.23       & 1.43       & 1.66       & 1.93       &            &            &            &            &             \\
                 & \textbf{KF+} & 0.93       & 1.22       & 1.40       & 1.61       & 1.91       & 0.10       & 0.30       & 0.21       & 0.07       & 0.26        \\ 
\hline
\textbf{UNEMP}   & \textbf{EWM} & 0.16       & 0.37       & 0.56       & 0.76       & 0.91       & 1.00 -     & 0.90       & 1.00 -     & 1.00 -     & 1.00 -      \\
                 & \textbf{KF}  & 0.15       & 0.36       & 0.55       & 0.74       & 0.89       & 0.94       & 0.69       & 1.00 -     & 1.00 -     & 0.98 -      \\
                 & \textbf{CWM} & 0.14       & 0.36       & 0.53       & 0.70       & 0.87       &            &            &            &            &             \\
                 & \textbf{KF+} & 0.14       & 0.36       & 0.54       & 0.72       & 0.87       & 0.23       & 0.30       & 0.88       & 0.94       & 0.57        \\ 
\hline
                 &              &            &            &            &            &            &            &            &            &            &             \\
\multicolumn{12}{l}{$+$ Significant at the five percent level \textbf{against} the CWM.}                                                                           \\
\multicolumn{12}{l}{$-$ Significant at the five percent level \textbf{for} the CWM.}                                                                              
\end{tabular}
\end{center}
\end{table}

Both the KF and the CWM easily beat the EWM on average and did so with greater clarity when the forecast horizons were longer. This result can be explained by the content of Figure 4, and supports prediction 6. Figure 4 shows that for longer forecast horizons, the diversity in $p$ increases, yet forecasters remain highly adept on average. According to Prediction 6, that situation is most detrimental for the EWM compared with the KF.

\begin{figure*}
\begin{center}
\caption{\em Median Values of Estimated $p$ across Forecast Horizons}
\centerline{\includegraphics[width=0.55\textwidth]{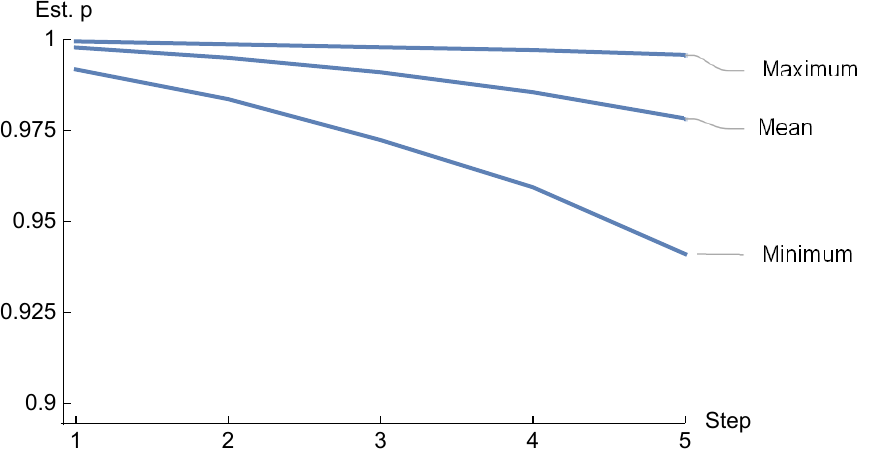}}
\label{fig:4}
\end{center}
\end{figure*}

Meanwhile, the KF beat the CWM on average for 1, 2, and 3 steps ahead, but was beaten on average by the CWM when the horizons were 4 and 5 steps ahead. This finding can again be explained by the content of Figure 4, and supports Prediction 8. Figure 4 shows that longer forecast horizons affect the distribution of $p$ not only by reducing the typical value of $p$ but by reducing the typical minimum by significantly more than it reduces the typical maximum. Consequently, the distribution of $p$ becomes increasingly defined by negative skewness caused by a tail that stretches farther into the region of lower $p$. In contrast, for shorter horizons, the general level of $p$ is not only higher, but the tail is also much shorter. Since Prediction 8 states the KF will beat the SR with a more significant margin when higher reliability defines the excluded estimates, the finding provides confirming evidence.  

The KF+, however, performed best of all, outperforming the KF and the CWM on average for every forecast horizon. Furthermore, just like its more inclusive relative, the KF+ outperformed with greater clarity on shorter forecast horizons. This again supports Prediction 8, since the effect of discounting all forecasters with negative CW is to shorten the tail of the $p$ distribution.

\subsection{Comparing Smaller, Wiser Crowds.}
\noindent Next, we examined the performance of the KF versus the CWM, and the KF against the EWM, as we discounted an increasing number of the lowest performing forecasters. As shown by Figure 5 and 6, this investigation gave greater nuance to the conclusion based on the entire data. In particular, as shown by Figure 5, although the EWM was inferior before, it now outperformed the KF on average when it considered only the ten-highest performing active forecasters based on $p$. This finding is consistent with Prediction 4, which states that the EWR will tend to beat the KF when the estimates to be combined have similar reliability. By definition, excluding the worst performers makes the included forecasters more alike. For this same reason, we can also explain why the EWM deteriorates so quickly compared with the KF when both include an increasing number of inferior forecasters. Quite simply, increasing the number of participants considered has the opposite effect on diversity.

\begin{figure*}
\begin{center}
\caption{\em The KF versus the EWM across Forecast Horizons and Subsets of Forecasters}
\centerline{\includegraphics[width=0.40\textwidth]{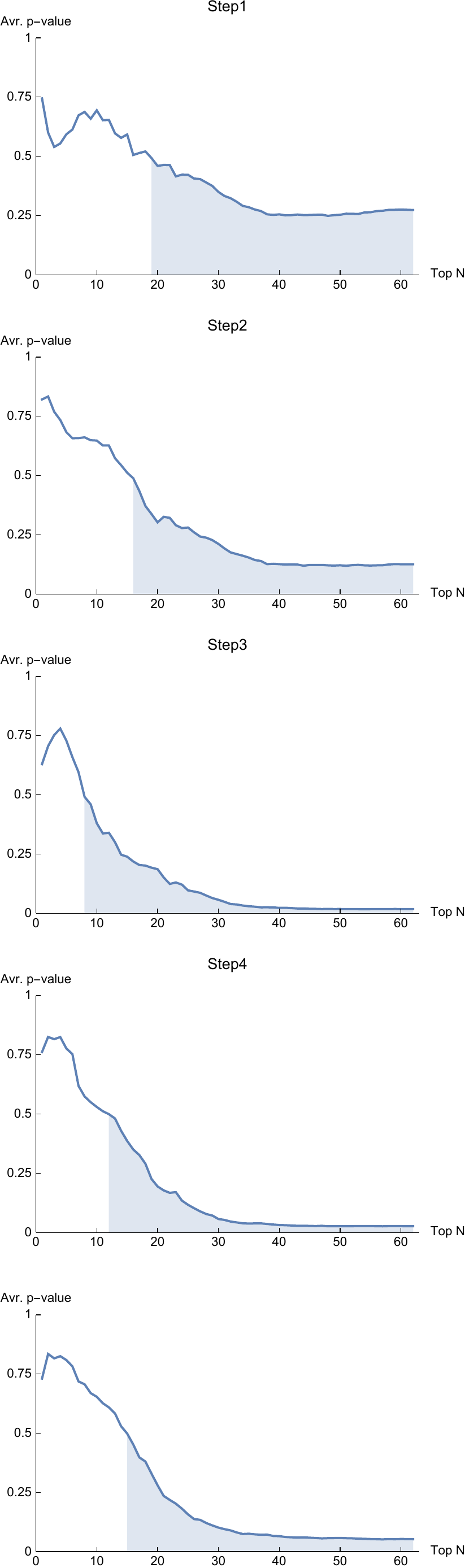}}
\label{fig:5}
\end{center}
\end{figure*}

Figure 6 compares both the KF and the KF+ with the CWM, uncovering three new patterns. First, the KF beat the CWM model more clearly as we discounted all but the top two active forecasters, and second, the point of greatest relative performance by the KF involved more individuals for shorter forecast horizons and fewer individuals for longer horizons. Finally, whenever the KF beat the CWM by the highest margin, it matched the performance of the KF+, except for the most extended forecast horizon where the KF+ performed noticeably better.

The first pattern supports Prediction 8 and Prediction 9 in unison. According to Prediction 8, the KF will beat the SR when estimates by the subset and by the other entity are both defined by high reliability. Moreover, Prediction 9 states the effect will be more significant when higher reliability defines both estimates. Since the increased focus on the best performers makes the forecasters more homogeneous and raises their average level of adeptness, we notice the support for these predictions quite easily. 

We can explain the second pattern by considering that more forecasters become more reliable (i.e., more forecasters have high values of $p$) when the forecasting horizon shortens (Figure 4). Consequently, the KF improves more quickly as it discounts the worst forecasters in surveys relating to shorter forecasts. 

Finally, we cannot fully explain the last observation. The idea that KF+ and KF consider precisely the same forecasters at that point where KF performs best is not entirely adequate. For that idea to be comprehensive, the gap observed in the case of the most extended forecast horizon should not be there. Accordingly, we cannot reject the idea that CW provides a way to find people with unique insight into long-scale developments, whose estimates should not be fused with those by people lacking such knowledge. What is clear, however, is that having perhaps identified such people, the CWM weighs their estimates less optimally than the KF.  

\begin{figure*}
\begin{center}
\caption{\em The KF versus the CWM across Forecast Horizons and Subsets of Forecasters}
\centerline{\includegraphics[width=0.40\textwidth]{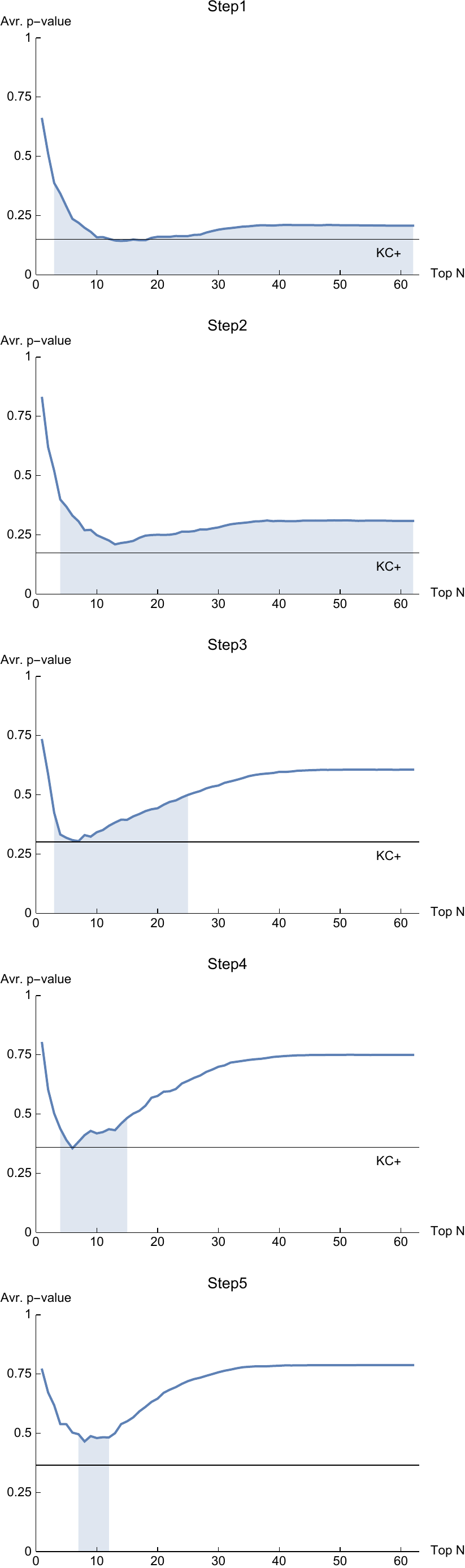}}
\label{fig:6}
\end{center}
\end{figure*}

\section{Discussion}
\noindent Unpredictable fluctuations and disturbances that are not part of the signal are a central feature of the nervous system \citep{Faisal2008}. That observation has led to the idea that our representations of the environment are probabilistic, which means that when we make a judgment, we are sampling from a probability distribution \citep{Ma2006}. In turn, this idea led to the theory \citep{Vul2008} that when we observe the wisdom of the crowd, what we are witness to is not a phenomenon working on a judgment distribution formed by people who are certain, and who would form the same distribution if asked to judge again. Instead, the wisdom of the crowd arises from each person having drawn an uncertain estimate from their probabilistic representation of the circumstances. 

But if the fundamental nature of our cognitive system causes the wisdom of crowds, then it creates some basic questions for researchers who work in the particular domain of management relating to better decision-making. If a managerial judgment is a random variable, perhaps we can successfully use methods from engineering that aim to estimate physical properties of the environment by filtering signals from the noise in multiple sensors. From the perspective of traditional management, the idea perhaps seems a little far fetched, but from other perspectives \citep{Luce1972}, which see the brain as a measuring device, the idea is quite reasonable. In any case, we aimed to find out in this paper, and our results offer positive indications.

What we did was move the Kalman Filter \citep{Kalman1960} (KF) from the domain of electrical engineering to management with the help of the AQ model of probabilistic judgment \citep{Nash2017}, which recently appeared in the field of mathematical psychology. Using that model, we matched the KF against the classic Equal-Weight Model (EWM) and the Contribution Weighted Model \citep{Budescu2015} (CWM), which has proven, on numerous occasions \citep{Budescu2015, Chen2016}, to be an excellent rule for combining estimates from many sources.

Our predictions based on the AQ suggest, and our empirical findings show, that using the KF is better than using the EWM or the CWM in those situations where the uncertainty of estimates is generally small, while there remains some diversity in the ability of people to judge. However, when the adeptness of judges is smaller, which it is, for example, when forecasting horizons are longer, then the CWM outperforms. The reason is that while the CWM works on a subset of more certain estimates, the KF includes all the available forecasts. 

That led to the idea that it should be possible to give the KF greater effect by using it on the specific subset of estimates identified by the CWM, and that proved correct. The KF+, as we called this version of KF, beat the CWM on average across all forecast horizons. Consequently, since the KF and the CWM were directly comparable in this situation, we may conclude, as Budescu and Chen \cite{Budescu2015} recently did, that the CWM works not so much because it applies differential weights optimally, but because it can identify forecasters, whose estimates should not be given any weight at all. In comparison, what we demonstrated in this paper is that when the CWM has identified the elite, then the KF can perform the task of differential weighting exceedingly well. 

In summary, we have shown not only that using the Kalman Filter on judgments has merit in theory, but also that in practice, this invention, which helped Armstrong safely to the moon and back \citep{McGee1985}, can help managers navigate through difficult problems by giving them stronger signals before they decide.


\bibliography{C:/Users/uwn/Dropbox/LaTeX/References/arXiv_Kalman}{}

\end{document}